\documentclass[iop,apj,tighten,twocolappendix,numberedappendix]{emulateapj}
\usepackage{apjfonts}

\usepackage{graphicx}
\usepackage{amsmath}
\usepackage{amssymb}
\usepackage{color}
\usepackage{mathtools}
\usepackage[breaklinks,colorlinks,citecolor=blue,linkcolor=red]{hyperref} 
\usepackage[all]{hypcap}

\newcounter{RomanNumber}

\newcommand{\degree}{^\circ}
\newcommand{\ue}{\mathrm{e}}

\begin{document}

\title{Tidal Disruption of a Main-Sequence Star by An Intermediate-mass Black Hole: \\
A bright decade}

\author{Jin-Hong Chen, Rong-Feng Shen}
\affil{School of Physics and Astronomy, Sun Yat-Sen University, Guangzhou, 510275, P. R. China; chenjh258@mail2.sysu.edu.cn, shenrf3@mail.sysu.edu.cn}

\begin{abstract}
There has been suggestive evidence of intermediate-mass black holes (IMBHs; $10^{3-5} M_\odot$) existing in some globular clusters (GCs) and dwarf galaxies, but IMBHs as a population still remain elusive. As a main-sequence (MS) star passes too close by an IMBH it might be tidally captured and disrupted. We study the long-term accretion and the observational consequence of such tidal disruption events. The disruption radius is hundreds to thousands of the BH's Schwarzschild radius, so the circularization of the falling-back debris stream is very inefficient due to weak general relativity effects. Due to this and a high mass fallback rate, the bound debris initially goes through a $\sim 10$ yr long super-Eddington accretion phase. The photospheric emission of the outflow ejected during this phase dominates the observable radiation and peaks in the UV/optical bands with a luminosity of $\sim 10^{42}\ {\rm erg\ s^{-1}}$. After the accretion rate drops below the Eddington rate, the bolometric luminosity follows the conventional $t^{-5/3}$ power-law decay, and X-rays from the inner accretion disk start to be seen. Modeling the newly reported IMBH tidal disruption event candidate 3XMM J2150-0551, we find a general consistency between the data and the predictions. The search for these luminous, long-term events in GCs and nearby dwarf galaxies could unveil the IMBH population.
\end{abstract}


\section{Introduction}
\label{introduction}
There exists much evidence for the existence of stellar-mass ($\sim 10 M_{\odot}$) and  supermassive ($\sim 10^6 - 10^9 M_{\odot}$) black holes (BHs), but still no firm evidence for the existence of intermediate-mass black holes (IMBHs; $\sim 10^{3-5} M_{\odot}$), which fill a gap between these mass ranges.

The centers of globular clusters (GCs) have long been suspected to harbour IMBHs \citep{Miller_Production_2002, Zwart_The_2002}. Discoveries of the central $10^3 - 10^5 M_{\odot}$ BHs have been made from spatially resolved kinematics with the \textit{Hubble Space Telescope} for the Galactic GC M15 \citep{Gerssen_2002} and the Andromeda GC G1 \citep{Gebhardt_A_2002}. However, \cite{Baumgardt_A_2003} found that G1's data can be fit equally well by a model without a central BH. Other GCs show marginal evidence of a central IMBH \citep{Noyola_Gemini_2008}. Some ultra-luminous X-ray sources (those with $L_X > 10^{39} erg \ s^{-1}$) are also thought to be accreting IMBH candidates. Other promising locations that may harbour IMBHs include the centers of dwarf galaxies \citep{Volonteri_The_2003, Moran_Black_2014, Baldassare_X_2017}

In a GC harbouring an IMBH, a star is occasionally perturbed into a highly eccentric orbit that causes it move too close to the IMBH and thus become tidally disrupted. Past attention has been focused on the disruption of a white dwarf by an IMBH \citep{Haas_TIDAL_2012, Rosswog_Atypical_2008}, due to the shorter time scale and thus the higher energy release power. This type of event has been used to explain some long gamma-ray bursts \citep{Lu_A_2008} and a very energetic tidal disruption event (TDE; \cite{Krolik_SWIFT_2011}). In this paper we consider a much more frequent encounter---the disruption of a main-sequence star by an IMBH. Our study is an extension of that by \cite{Ramirez-Ruiz_THE_2009} who studied the disruption process using smoothed-particle hydrodynamics simulation and briefly discussed the observability of IMBH TDEs. We focus on the long-term evolution of the accretion and the radiation properties.

In Section 2, we describe the general  process of tidal disruption of a main-sequence star by an IMBH, and the subsequent debris fallback. In Section 3, we discuss a long-term super-Eddington accretion phase based on the inefficient circularization and calculate the bolometric light curve and temperature accordingly. In Section 4, we estimate the detection rate by current optical surveys such as the Zwicky Transient Factory (ZTF) and X-ray observatories such as \textit{Chandra}. We summarize and discuss the results in Section 5.

\section{Tidal Disruption process}
A TDE occurs when the flying-by star's pericenter radius $R_{\rm p}$ reaches the tidal radius $R_{\rm T} = R_*(M_{\rm h}/M_*)^{1/3}$ \citep{Rees_Tidal_1988,Phinney_Manifestations_1989}. Here $M_{\rm h} \equiv {M_4} \times 10^4 M_{\odot}$, $R_* \equiv r_* \times R_{\odot}$, $M_* \equiv m_* \times M_{\odot}$ are the BH's mass and the star's radius and mass, respectively. The penetration factor is defined as $\beta \equiv R_{\rm T} / R_{\rm p}$. In units of the BH's Schwarzschild radius $R_{\rm S} = 2GM_{\rm h}/c^2$, the pericenter radius is \begin{equation}
R_{\rm p} \simeq 500\ \beta^{-1} M_4^{-2/3} r_* m_*^{-1/3}\ R_{\rm S}.
\end{equation}

When the star is tidally disrupted by an IMBH, the debris would have a range in specific energy of $\sim \pm GM_{\rm h} R_*/R_{\rm T}^2$ \citep{Lacy_The_1982, Li_The_2002} due to its location in the IMBH's potential well. The debris closer to the BH has negative specific energy, and that farther away has positive specific energy. So approximately half of the star's (the closer part of the debris) is bound to the IMBH and the rest flies away on hyperbolic orbits with escape velocity. The most closely bound debris with the specific energy $E_{\rm mb} \simeq -GM_{\rm h} R_*/R_{\rm T}^2 \simeq -GM_{\rm h}/(2a_{\rm mb})$ is the first to return to the pericenter, here $a_{\rm mb} \simeq R_{\rm T}^2/(2R_*) \simeq 5.7 \times 10^3\ M_4^{2/3} r_* m_*^{-2/3}\ R_{\rm S}$ is the semi-major axis of the orbit of the most closely bound debris. The eccentricity of this orbit is $e_{\rm mb} = 1 - R_{\rm p}/a_{\rm mb} \simeq 1-0.09\ \beta^{-1} M_4^{-1/3} m_*^{1/3}$. Its period of this orbit 
\begin{equation}
P_{\rm mb} = 2 \pi \sqrt{a_{\rm mb}^3/GM_{\rm h}} \simeq 4.4\ M_4^{1/2} r_*^{3/2} m_*^{-1}\ {\rm day}
\label{mbperiod}
\end{equation}
determines the characteristic timescale of the debris fallback.

\begin{figure}
\centering
 \includegraphics[scale=0.1]{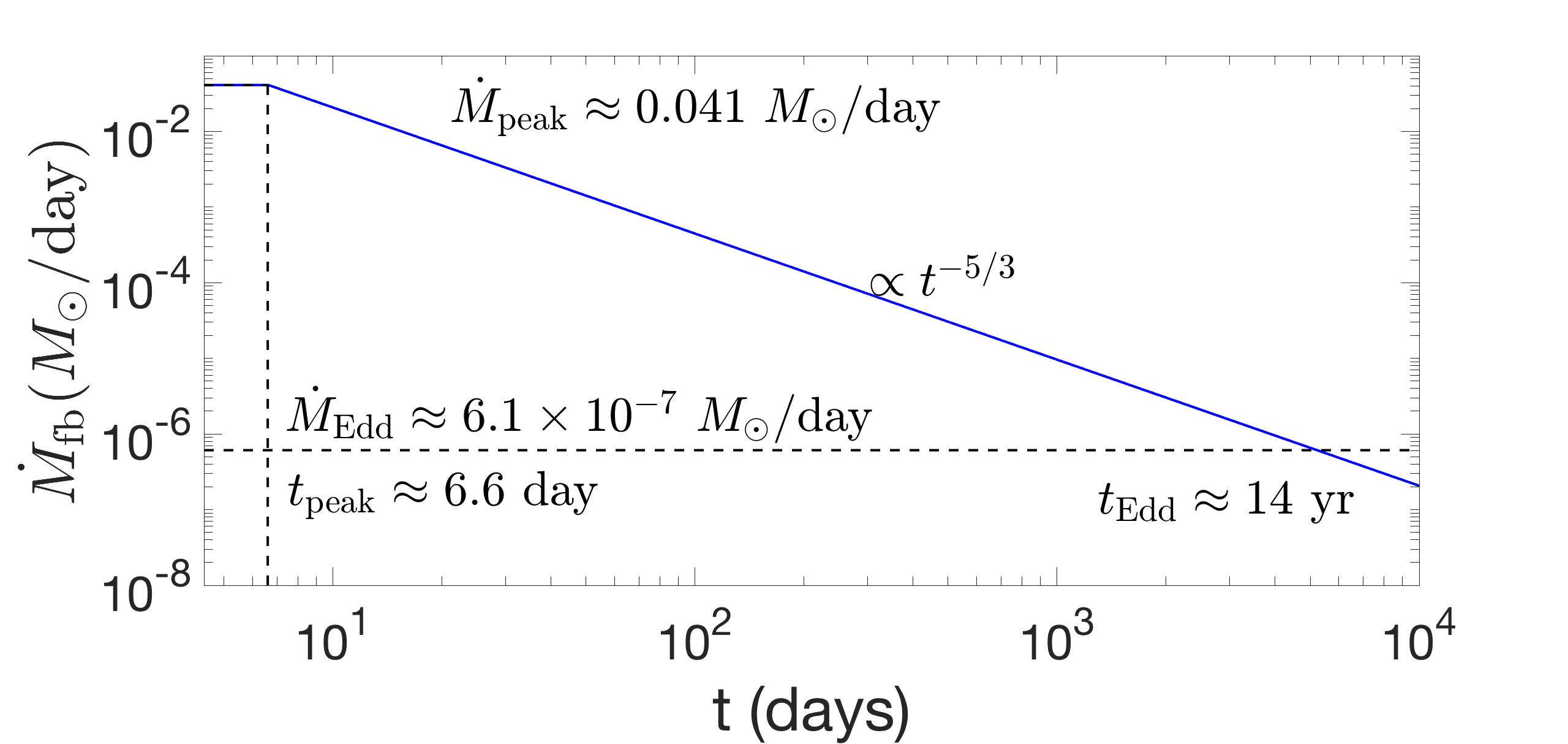}
\caption{Simplified mass fallback rate history for the disruption of a Sun-like star ($m_*=r_*=1$) by a $10^4M_{\odot}$ IMBH. After $P_{\rm mb} \sim 4.4\ {\rm days}$, the debris begins to arrive back at the pericenter. After $t_{\rm peak} \sim 6.6\ {\rm days}$, the mass fallback rate follows the fiducial decay rate $t^{-5/3}$. Here we neglect the ascending stage between $P_{\rm mb}$ and $t_{\rm peak}$, and let $\dot M_{\rm fb} \sim \dot M_{\rm peak}$. }
\label{fallback_rate}
\end{figure}

The less bound debris follows the most bound debris in returning, at a rate that drops with time as \citep{Rees_Tidal_1988, Phinney_Manifestations_1989,Lodato_Stellar_2009,Ramirez-Ruiz_THE_2009,Guillochon_HYDRODYNAMICAL_2013}:
\begin{equation}
\dot M_{\rm fb} \simeq \dot M_{\rm peak} (\frac{t}{P_{\rm mb}})^{-5/3}.
\label{fallback_eq}
\end{equation}
We approximate the overall fallback rate history as follows. The mass fallback rate reaches its peak value at the time $t_{\rm peak} = 1.5P_{\rm mb}$ \citep{Evans_The_1989}; between $P_{\rm mb}$ and $t_{\rm peak}$, the fallback rate remains constant at $\dot M_{\rm peak}$, and after $t_{\rm peak}$ it starts to decay as $t^{-5/3}$, as illustrated in Figure~\ref{fallback_rate}. For a complete disruption case, the total mass of fallback debris will be $\sim M_*/2$, therefore, $\dot M_{\rm peak} \simeq 0.2 M_*/P_{\rm mb}$.

\section{Long-term super-Eddington accretion phase}
\label{long}

Initially, the fallback rate $\dot M_{\rm peak} \simeq 6.7 \times 10^4\ \eta_{-1} M_4^{-3/2} r_*^{-3/2} m_*^2\ \dot M_{\rm Edd}$ is far greater than the Eddington rate $\dot M_{\rm Edd} \equiv L_{\rm Edd}/(\eta c^2)$. Here $\eta \simeq 0.1$ is the assumed efficiency of converting accretion power to luminosity. From Eq. (\ref{fallback_eq}) the time when the fallback rate drops to $\dot M_{\rm Edd}$ is
\begin{equation}
t_{\rm Edd} \simeq 14\ \eta _{0.1}^{3/5} M_4^{-2/5} r_*^{3/5} m_*^{1/5}\ {\rm yr}.
\label{eddingtontime}
\end{equation}

During the super-Eddington phase, $t<t_{\rm Edd}$, various energy dissipation processes might produce winds or outflows \citep{Strubbe_Optical_2009, Lodato_Multiband_2011, Metzger_A_2016}. If the accretion time scale $\tau_{\rm acc}$ is shorter than the orbital period of the debris, this debris could be accreted rapidly, and the accretion rate $\dot M_{\rm acc} (t)$ should follow the fallback rate $\dot M_{\rm fb} (t)$. However, this might not occur until the stream settles into a circular disk. We discuss below the process of circularization. 

The orbital angular momentum of the star flying by the tidal radius should be almost conserved. In order for the debris to circularize at the circularization radius $R_{\rm c}=2R_{\rm p}$, the debris has to lose a specific amount of orbital energy of the order
\begin{equation}
E_c =  \frac{GM_{\rm h}}{2R_{\rm c}} \sim 2.2 \times 10^{17}\ \beta M_4^{2/3} r_*^{-1} m_*^{1/3}\ {\rm erg/g}.
\label{circularization_energy}
\end{equation} 
Theoretical analysis \citep{Piran_DISK_2015, Svirski_Elliptical_2017} and numerical simulations \citep{Shiokawa_GENERAL_2015, Bonnerot_Disc_2016} have shown that the process of circularization might be very ineffective\footnote{The efficiency of circularization might be enhanced if the pericenter radius is very close to the Schwarzschild radius \citep{Hayasaki_Finite_2013}, but we consider the case of $\beta \sim 1$ in this paper, since it is the most likely situation.}. In the Appendix, we analyze the efficiency of circularization by considering several main effects: periastron nozzle shock, apsidal intersection, thermal viscous shear, and magnetic shear, and find that all of them are very inefficient with regard to circularization. Therefore, we can reasonably expect that tidal disruption of a main-sequence star by an IMBH will undergo a very slow circularization.

\begin{figure}
\centering
 \includegraphics[scale=0.1]{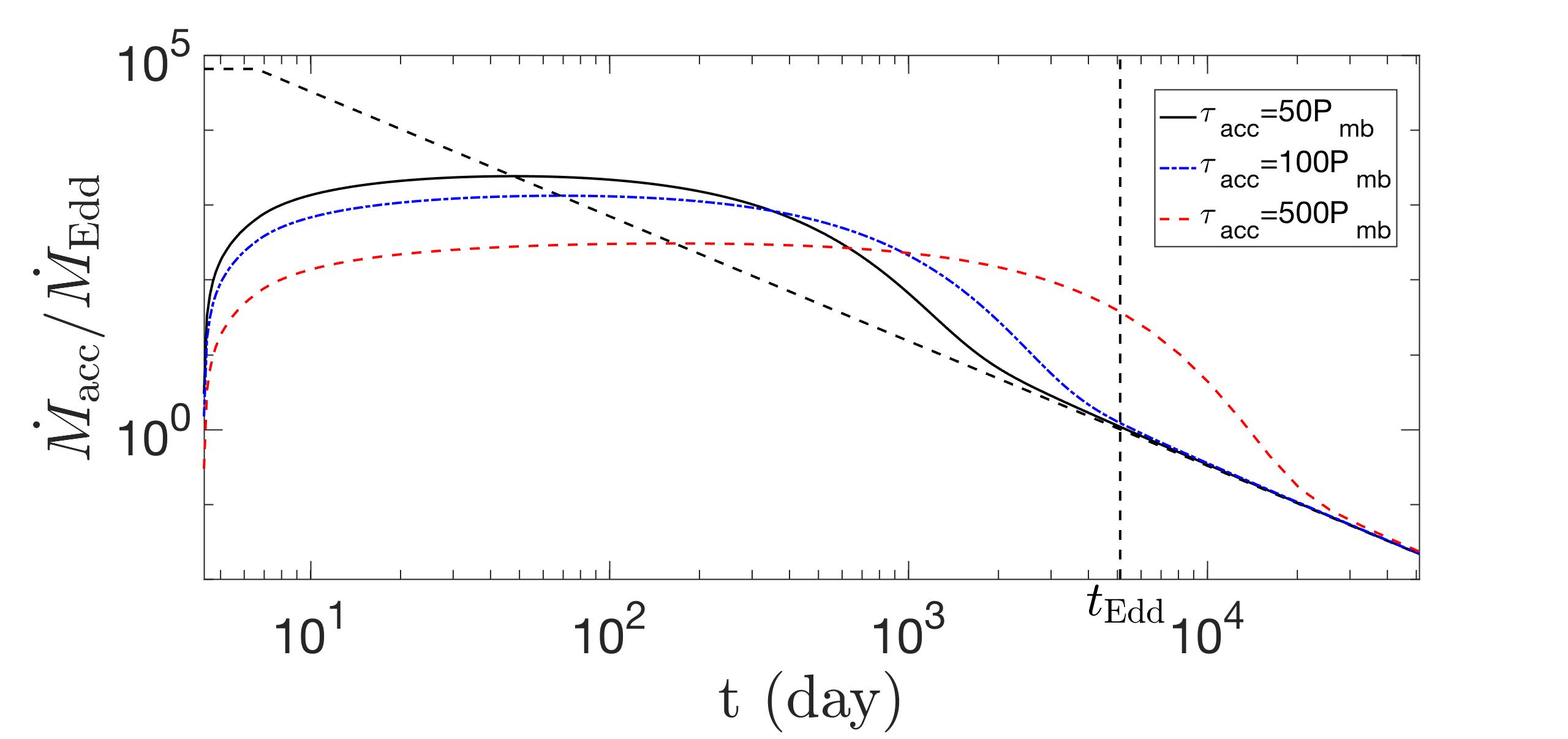}
\caption{Disk accretion rate history of a Sun-like star having been tidally disrupted by a $10^4 M_{\odot}$ IMBH ($\beta = 1$). The accretion rate is slowed relative to the fallback rate (the black dashed line) by the accretion timescale $\tau_{\rm acc}$. The black dashed line corresponds to the fallback rate (see Figure \ref{fallback_rate}). The black solid line, blue dot-dashed line, and red dashed line correspond to the accretion rate with accretion timescales $\tau_{\rm acc} = 50P_{\rm mb}$, $100P_{\rm mb}$ and $500P_{\rm mb}$, respectively.}
\label{accretion_rate}
\end{figure}

Since the exact details of how the streams evolve and form a disk are still unclear, especially in the super-Eddington phase, we adopt the following function to demonstrate the relation between the slowed accretion rate $\dot M_{\rm acc}(t)$ and the fallback rate \citep{Kumar_Mass_2008,Lin_A_2017,Mockler_Weighing_2018}:
\begin{equation}
\dot M_{\rm acc}(t)=\frac{1}{\tau_{\rm acc}} \left(\ue^{-t/ \tau_{\rm acc}}\int_{P_{\rm mb}}^t \ue^{t'/ \tau_{\rm acc}} \dot M_{\rm fb}(t')\ dt' \right),
\label{accretion_rate_eq}
\end{equation}
where the constant $\tau_{\rm acc}$ represents the `slowed' accretion time scale. Figure \ref{accretion_rate} shows the history of $\dot M_{\rm acc}(t)$ for $\tau_{\rm acc} = 50P_{\rm mb}, 100P_{\rm mb},\ {\rm and}\ 500P_{\rm mb}$. The accretion rate first rises rapidly, on a timescale $\sim P_{\rm mb}$ (which corresponds to the duration of the peak mass supply from the fallback) to a persisting plateau phase lasting for a period of $\sim \tau_{\rm acc}$, then settles down to the decaying $\dot M_{\rm fb}(t)$ curve.

During the super-Eddington accretion phase ($\dot M_{\rm acc} \gtrsim \dot M_{\rm Edd}$), the existence of wind (outflow) should act as a way to regulate the luminosity \citep{Krolik_JETS_2012, Piran_Jet_2015}, therefore, we follow the simplified approach of \cite{King_Black_2016} and \cite{Lin_A_2017} to let the luminosity scale logarithmically with $\dot M_{\rm acc}$. When $\dot M_{\rm acc} \lesssim \dot M_{\rm Edd}$, the wind (outflow) should cease, thus $L \propto \dot M_{\rm acc}$. So overall, 
\begin{equation} 
L= \begin{cases}
[1+\log_{10}{({\dot M_{\rm acc}/ \dot M_{\rm Edd}})}]L_{\rm Edd},&\quad{\dot M_{\rm acc} \gtrsim \dot M_{\rm Edd}} \\
(\dot M_{\rm acc}/ \dot M_{\rm Edd}) L_{\rm Edd},&\quad{\dot M_{\rm acc} \lesssim \dot M_{\rm Edd}}
\end{cases}
\label{luminosity_eq}
\end{equation}

\begin{figure}
\centering
 \includegraphics[scale=0.1]{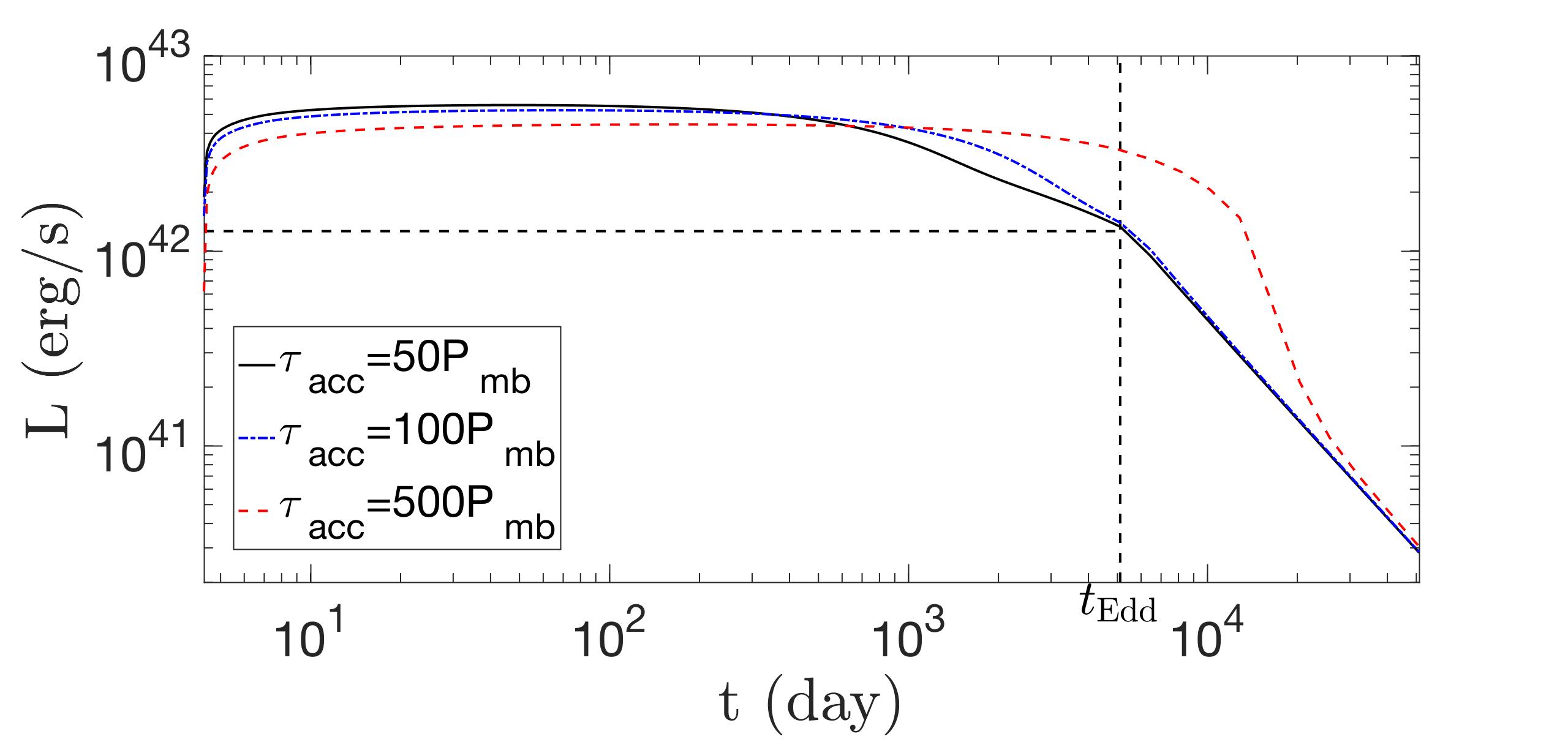}
\caption{Radiative luminosity history of a Sun-like star having been tidally disrupted by a $10^4 M_{\odot}$ IMBH ($\beta = 1$). Regulation of the luminosity (Eq. \ref{luminosity_eq}) due to disk wind loss is adopted for the super-Eddington phase ($\dot M_{\rm acc} \gtrsim \dot M_{\rm Edd}$), thereafter the wind stops and the luminosity follows the accretion rate, whose evolution has the familiar power law $t^{-5/3}$. The black solid line, blue dot-dashed line and red dashed line correspond to the luminosity with the accretion timescales $50P_{\rm mb}$, $100P_{\rm mb}$ and $500P_{\rm mb}$, respectively.}
\label{luminosity}
\end{figure}

Figure \ref{luminosity} shows the radiative luminosity history for $\tau_{\rm acc} = 50P_{\rm mb}, 100P_{\rm mb}\ {\rm and}\ 500P_{\rm mb}$, respectively. We see from Figure \ref{luminosity} that the event will radiate luminously for more than 10 yr as the bound debris is being accreted by the IMBH, and the luminosity is close to the Eddington luminosity. This result roughly agrees with the general anticipation of \cite{Ramirez-Ruiz_THE_2009}. Furthermore, if the accretion is slow enough, i.e., $\tau_{\rm acc} \gtrsim t_{\rm Edd}$, as is shown by the red dashed line in Figure \ref{luminosity}, this super-Eddington phase will last for much longer than $t_{\rm Edd}$.

Recent simulations of super-Eddington accretion disks \citep{Sadowski_Photon_2015,Sadowski_Three_2016,Jiang_Super_2017} suggest that their luminosity could exceed $L_{\rm Edd}$ by a factor that is much larger than the logarithm of $\dot M_{\rm acc}/ \dot M_{\rm Edd}$. However, these simulations assume relatively strong magnetic fields of $\sim 10^3$ G. In this paper, we consider the disruption of a Sun-like star, which has a weak magnetic field of $\sim 1$ G. According to magnetic flux conservation, $B \propto r^{-2}$, the strength drops as the debris expands by pressure or shocks. When the debris expands to radii $R_{\rm c} \simeq 43\ R_{\odot}$, the strength is expected to drop to $5 \times 10^{-4}$ G. Such a weak magnetic field falls far short of that required in those simulations which observed super-Eddington radiative luminosities. \cite{Bonnerot_Magnetic_2017} and \cite{Guillochon_Simulations_2017} studied the magnetic field evolution in TDEs. They found that the magnetic field evolves with the debris structure and the surviving core can enhance the field via the dynamo process. However, this process does not apply in the case of full disruption that we consider here. 

It has long been suspected that magnetorotational instability (MRI) could amplify the seed field in accretion disks \citep{Balbus_A_1991,Hawley_A_1991}. \cite{Hawley_A_1991} and \cite{Stone_Three_1996} showed that the magnetic field energy could rapidly increase from its initial value by about one order of magnitude during the MRI's nonlinear growth. However, such an enhancement of the field is still weaker than that required in those simulations observing supper-Eddington radiative luminosities. Furthermore, for eccentric disks which likely form in TDEs, how large the enhancement of the field will be is still unclear. Further discussions on magnetic field are given in Appendix \ref{magnetic}.

\begin{figure}
\centering
 \includegraphics[scale=0.1]{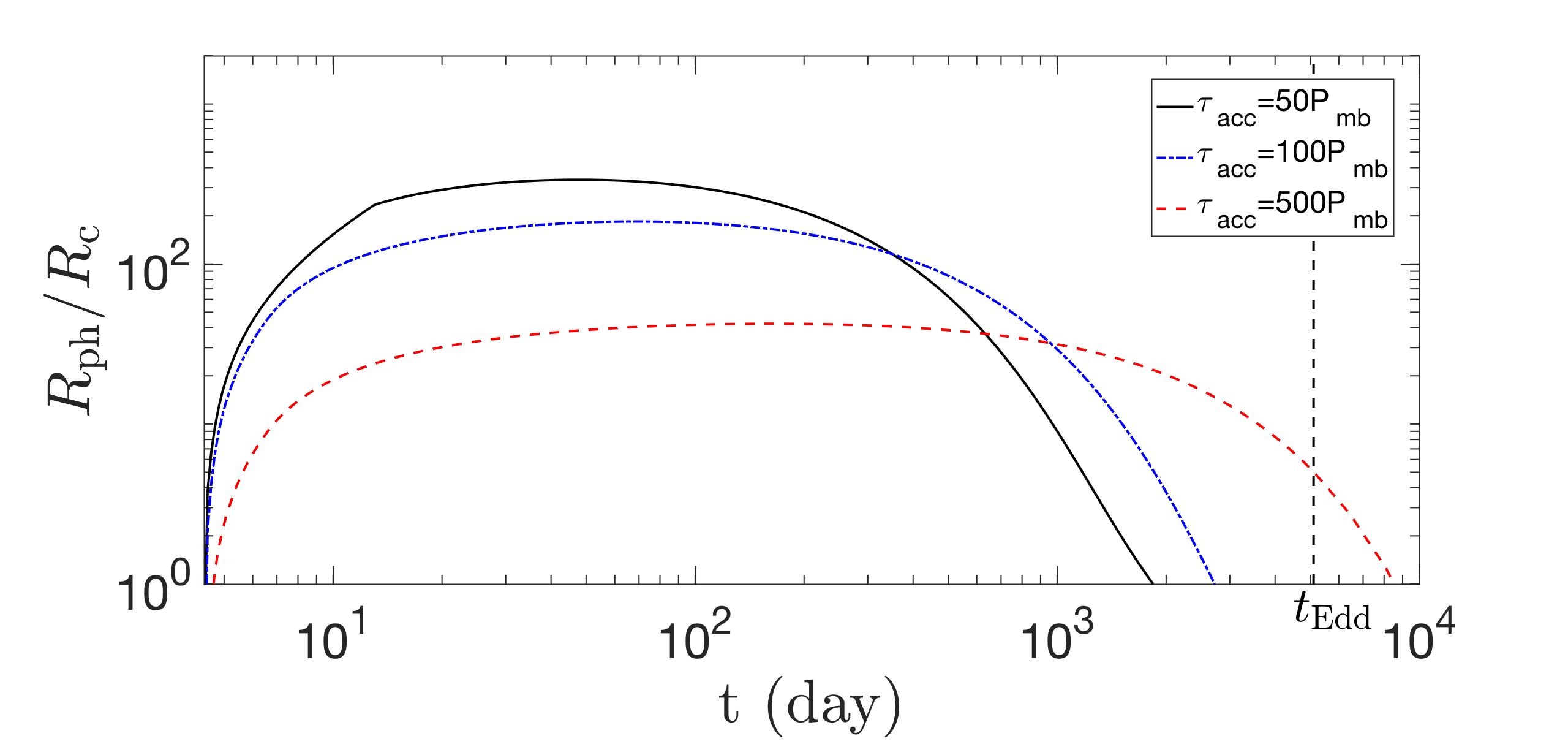}
\caption{Evolution of the electron scattering photosphere radius normalized by the circularization radius after the disruption of a Sun-like star by a $10^4 M_{\odot}$ IMBH ($\beta = 1$) for $f_{\rm w} = 1$.}
\label{photosphere}
\end{figure}

Next we discuss the spectral property of the radiation. We assume that the outflow is launched from $R_{\rm c}$ with the terminal velocity $v_{\rm w} \simeq f_{\rm w} (2GM_{\rm h}/R_{\rm c})^{1/2}$, scaling with the local escape velocity by a factor $f_{\rm w} \sim 1$. The outflow mass rate is assumed to be $\dot M_{\rm w} \simeq \dot M_{\rm acc} - \dot M_{\rm Edd}$, and thus the density profile is $\rho(R) = \dot M_{\rm w}/(4\pi R^2 v_{\rm w})$.
We further assume that the source radiates from the quasi-sperical photosphere whose radius is given by $\int_{R_{\rm ph}} \kappa_{\rm es} \rho(R)\ dR=1$, where $\kappa_{\rm es} \simeq 0.34~{\rm cm^2\ g^{-1}}$ is the opacity for electron scattering for a typical gas composition. The Kramers opacity, including bound-free and free-free, is found to be much smaller than $\kappa_{\rm es}$ for the density and temperature appropriate to the current situation. Therefore, $R_{\rm ph} \simeq (\kappa_{\rm es} \dot M_{\rm w})/(4\pi v_{\rm w})$ or
\begin{equation}
\label{Rph}
\frac{R_{\rm ph}}{R_{\rm c}} \simeq 5~\eta_{-1}^{-1} f_{\rm w}^{-1} \left(\frac{R_{\rm S}}{R_{\rm c}}\right)^{1/2} \frac{\dot M_{\rm w}}{\dot M_{\rm Edd}}.
\end{equation}
Figure \ref{photosphere} shows the time evolution of the photosphere.

\begin{figure}
\centering
 \includegraphics[scale=0.1]{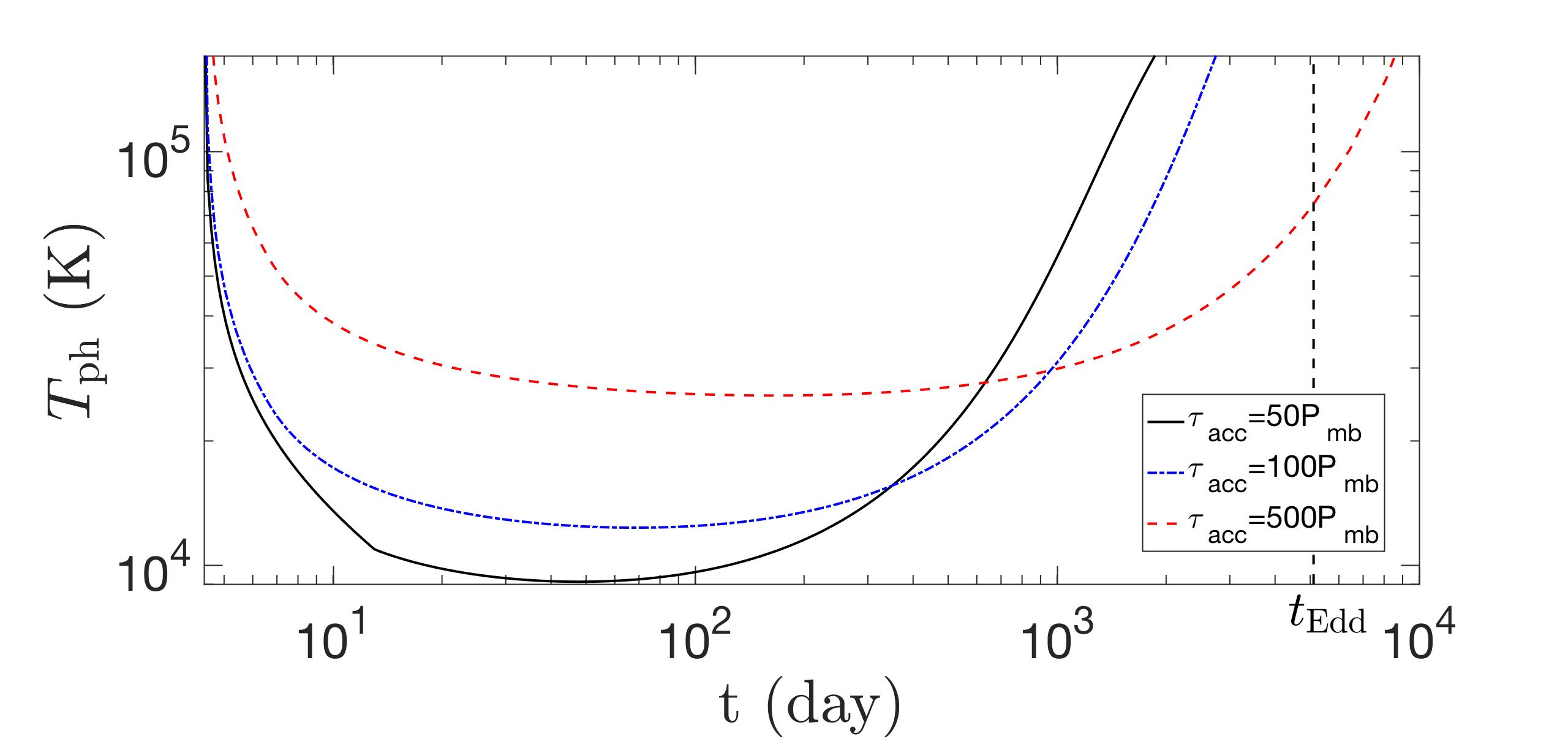}
\caption{Evolution of the photospheric temperature of the outflow during the super-Eddington phase of a Sun-like star tidally disrupted by a $10^4 M_{\odot}$ IMBH ($\beta = 1$) for $f_{\rm w} = 1$.}
\label{Tph}
\end{figure}

The general shape of the spectrum should be thermal, and a crude estimate of the characteristic temperature would be the effective temperature at the photosphere: $T_{\rm ph} \simeq [L/(4\pi R_{\rm ph}^2 \sigma)]^{1/4}$, or
\begin{equation}
T_{\rm ph} \simeq 3 \times 10^5\ \eta_{-1}^{1/2} f_{\rm w}^{1/2} M_4^{-1/4} R_{\rm c,3}^{-1/4} \times \left(\frac{L}{L_{\rm Edd}}\right)^{1/4} \left(\frac{\dot M_{\rm w}}{\dot M_{\rm Edd}}\right)^{-1/2}\ {\rm K},
\end{equation}
where $R_{\rm c,3} \equiv R_{\rm c}/(10^3~R_{\rm S})$. The temporal evolution of $T_{\rm ph}$ is plotted in Figure \ref{Tph}, which shows that a slowed accretion produces a much lower outflow rate; thus the photosphere is closer to the central source and lasts for a much longer time with a higher effective temperature. Basically, the spectrum peaks in the UV band until the photosphere recedes to $R_{\rm c}$ and then it peaks in the far UV and soft X-ray.

\section{Event Rate}

\begin{figure}
\centering
\begin{minipage}{20.5cm}
 \includegraphics[scale=0.1]{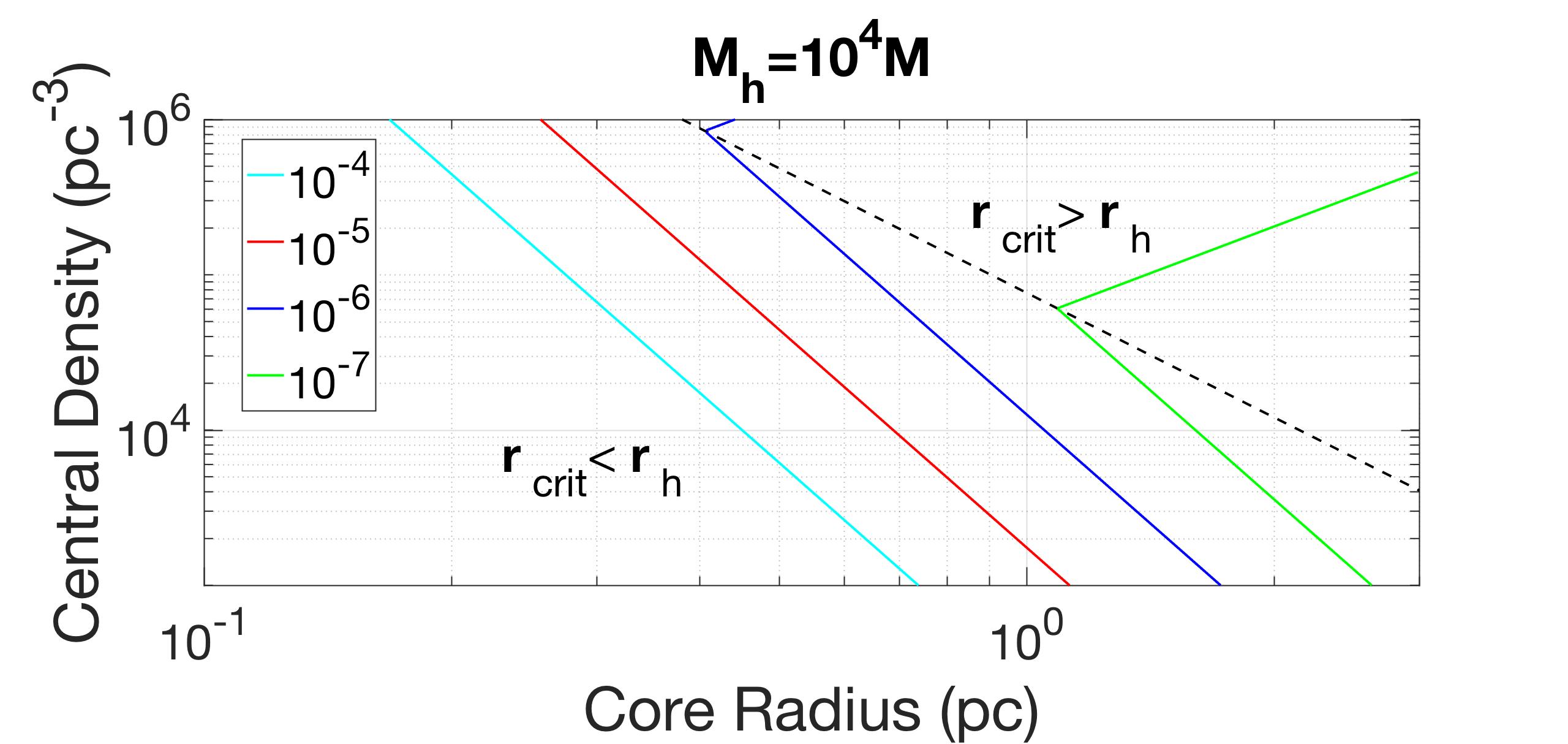}
 \end{minipage}
 \begin{minipage}{20.5cm}
 \includegraphics[scale=0.1]{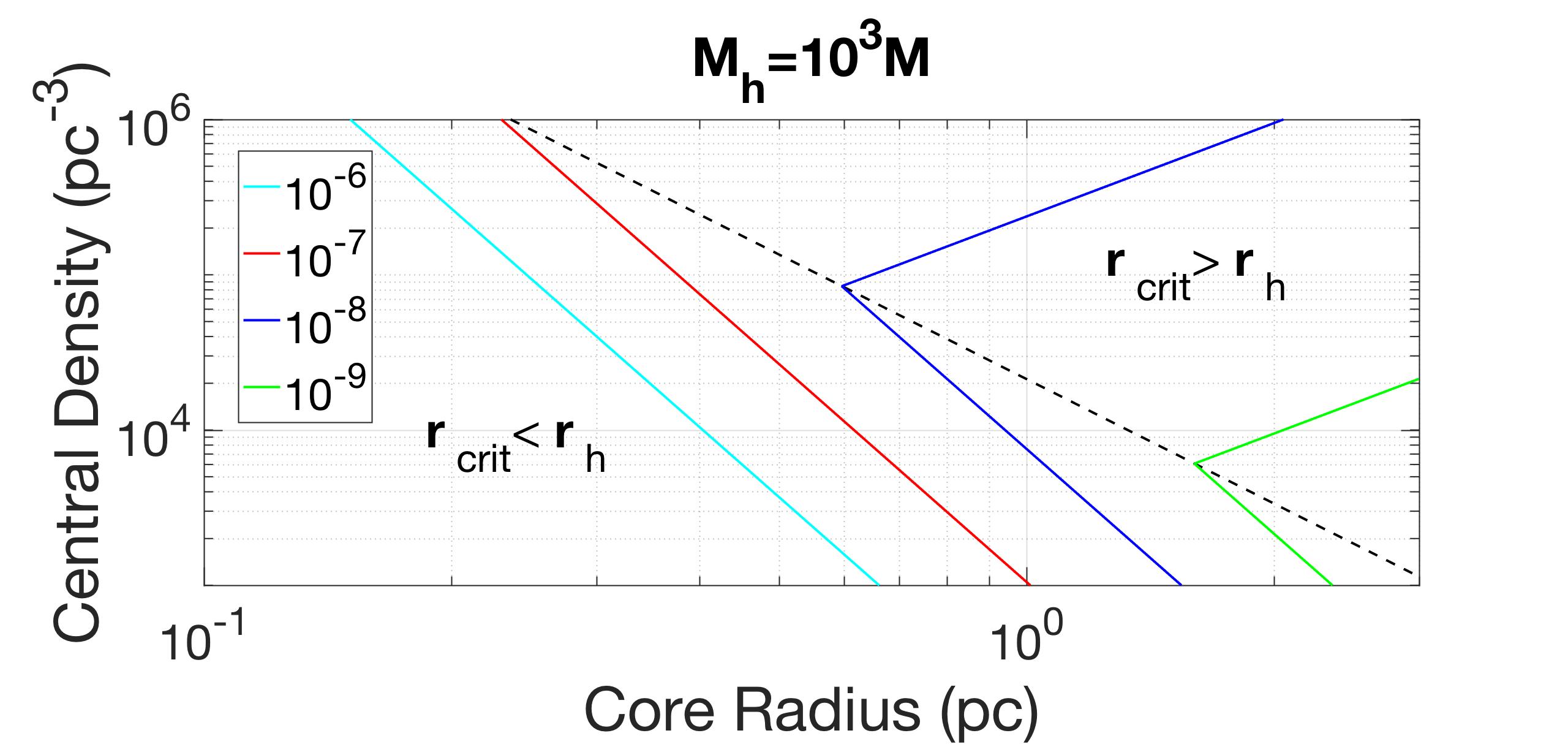}
 \end{minipage}
\caption{Event rate of stellar disruption by a $10^4M_{\odot}$ (top) and $10^3M_{\odot}$ (bottom) IMBH. The dashed black line divides the parameter space into zones of $r_{\rm crit} \lesssim r_{\rm h}$ and $r_{\rm crit} \gtrsim r_{\rm h}$. The colored lines represent different event rates (Eq. \ref{event_rate_eq}). For example, the event rate in a typical GC ($n_c = 10^6\ {\rm pc^{-3}}$, $R_{\rm c} = 1\ {\rm pc}$) is about $10^{-7} - 10^{-6}\ {\rm yr^{-1}}$ with $10^4M_{\odot}$ mass IMBH.}
\label{eventrate}
\end{figure}

The stellar tidal capture rate by a BH of $10^3 - 10^5M_{\odot}$ in a globular cluster was calculated by \cite{Frank_Effects_1976}, who gave the event rate
\begin{equation} 
ER= \begin{cases}
6.1 \times 10^{-9}\ M_4^{61/27} n_{\rm c,6}^{-7/6} R_{\rm c}^{-49/9}\ {\rm yr^{-1}},&\quad{r_{\rm crit} \lesssim r_{\rm h}}, \\
4.4 \times 10^{-7}\ M_4^{4/3} n_{\rm c,6}^{1/2} R_{\rm c}^{-1}\ {\rm yr^{-1}},&\quad{r_{\rm crit} \gtrsim r_{\rm h}}.
\end{cases}
\label{event_rate_eq}
\end{equation}
Here $R_{\rm c}$ is the core radius of the cluster in units of pc, $n_{\rm c,6} \equiv n_c/(10^6$ pc$^{-3})$ is the number density of stars in the core, and $r_{\rm h} \simeq GM_{\rm h}/v_{\rm c}^2$ ($v_{\rm c}$ is the velocity dispersion in the core) is the radius within which the gravitational potential is dominated by the BH alone. The critical radius $r_{\rm crit}$ is defined such that most stars on orbits with $r \lesssim r_{\rm crit}$ diffuse into low angular momentum `loss-cone' orbits within the `reference time' $t_{\rm R}$:
\begin{equation} 
r_{\rm crit}= \begin{cases}
30.6~M_4^{-20/27} n_{\rm c,6}^{4/3} R_{\rm c}^{32/9} r_*^{4/9} m_*^{-4/27}\ r_{\rm h},&\quad{r_{\rm crit} \lesssim r_{\rm h}} \\
13.1~M_4^{-5/9} n_{\rm c,6} R_{\rm c}^{8/3} r_*^{1/3} m_*^{-1/9}\ r_{\rm h},&\quad{r_{\rm crit} \gtrsim r_{\rm h}}
\end{cases}
\end{equation}

Eq. (\ref{event_rate_eq}) is plotted in Figure \ref{eventrate}. For a $10^4 M_{\odot}$ IMBH, and for $R_{\rm c} \sim 1~pc$ and $n_{\rm c,6} \sim 1$ in a typical GC \citep{Harris_A_1996}, we have $r_{\rm crit} > r_{\rm h}$, so the event rate of stellar disruption is about  $10^{-7}\ M_4^{4/3} n_{c, 6}^{1/2} R_{\rm c}^{-1}\ {\rm yr^{-1}}$. For one GC, this event rate is very small. One needs to sample a relatively large volume within our nearby universe. We estimate the detection rate below.

The Local Supercluster has a diameter of 33 Mpc, and contains $\sim 100$ groups and clusters of galaxies, and therefore a total of $\sim 50,000$ galaxies. So the number density of galaxies is $\sim 1\ {\rm Mpc^{-3}}$. Taking the number of GCs per galaxy to be 100 \citep{Harris_A_1996}, the GC space density is estimated to be $n_{\rm GC} \sim 100\ {\rm Mpc^{-3}}$. Below we will assume that every GC  harbours an IMBH, so it is a crude estimation. 

The ZTF is a new time-domain survey with the R-band limiting magnitude $m_R \simeq 20.5$ and its field of view is more than 3750 square degrees per hour \footnote{\url{http://www.ztf.caltech.edu}}. At $\nu = 4.5 \times 10^{14}\ {\rm Hz}$, the corresponding spectral flux density limit is given by $m_R \simeq -2.5\ \lg{(f_{\nu}/3631J_y)}$, thus $f_{\nu} \simeq 2 \times 10^{-28}\ {\rm erg\ s^{-1}\ Hz^{-1}\ cm^{-2}}$. 
The limiting distance of the ZTF detection horizon for an IMBH TDE is
\begin{equation}
d_{\rm lim} = \left[\frac{L_{\nu}(T_{\rm ph})}{4 \pi f_{\nu}}\right]^{1/2} \simeq 170\ {\rm Mpc},
\end{equation}
where we take a blackbody spectrum of $T_{\rm ph} = 5 \times 10^4\ {\rm K}$ and a bolometric luminosity of $L = 5 \times 10^{42}\ {\rm erg\ s^{-1}}$. The ratio of field of view to the all-sky is $\sim 0.1$. So the average event rate in the observable volume is about  $0.1 \times 4 \pi d_{\rm lim}^3/3 \times n_{\rm GC} \times 10^{-7} \sim 20\ {\rm yr^{-1}}$, thus the detection rate by ZTF is $\sim 20\ {\rm yr^{-1}}$.

As accretion rate drops below the Eddington limit, we could observe more X-rays emitted from the central source, i.e., the inner accretion disk. We will assume it is a blackbody spectrum with an effective temperature $T_{\rm c} \sim 10^5\ {\rm K}$ at $R_{\rm c}$. The sensitivity of Chandra X-ray Observatory in the energy range of $0.2 - 10\ {\rm KeV}$ is $F_{\rm lim} \sim 4 \times 10^{-15}\ {\rm erg\ cm^{-2}\ sec^{-1}}$ in $10^5\ {\rm s}$, with a field of view of $1.0$ degree diameter \footnote{http://chandra.harvard.edu}. Thus, the ratio of field of view to the whole sky is $\sim  (0.5 \times \pi /180)^2/4 \sim 2 \times 10^{-5}$. The limiting distance of the Chandra detection horizon is 
\begin{equation}
d_{\rm lim, X} \simeq \left(\frac{L_{\rm 0.2-10KeV}}{4\pi F_{\rm lim}}\right)^{1/2} \simeq 150\ {\rm Mpc},
\end{equation}
where $L_{\rm 0.2-10KeV} = 4 \pi R_{\rm c}^2 \int_{0.2\ {\rm KeV}}^{10\ {\rm KeV}} \pi B_{\nu}(T_{\rm c})\ d\nu$. So the average event rate in the observable volume is about $2 \times 10^{-5} \times 4\pi d_{\rm lim, X}^3/3 \times n_{\rm GC} \times 10^{-7} \sim 0.003\ {\rm yr^{-1}}$. Since the X-ray emission of such an event lasts for $\gtrsim 10\ {\rm yr}$, the average chance of detection in the archival data of Chandra should be $\sim 0.03\ {\rm yr^{-1}}$.

After the acceptance of this paper for publication, we noticed that \cite{Fragione_Tidal_2018} estimated the TDE rate in GCs by semi-analytically calculating the evolution of the GC population in a host galaxy over cosmic time. They obtained $\sim 10^{-4} - 10^{-3}\ {\rm yr}^{-1}$ per galaxy, which is higher than what we estimate here ($10^{-5} - 10^{-4}\ {\rm yr}^{-1}$); the corresponding detection rate will then be higher.

\section{Application to 3XMM J2150-0551}
\begin{figure}
\centering
 \includegraphics[scale=0.1]{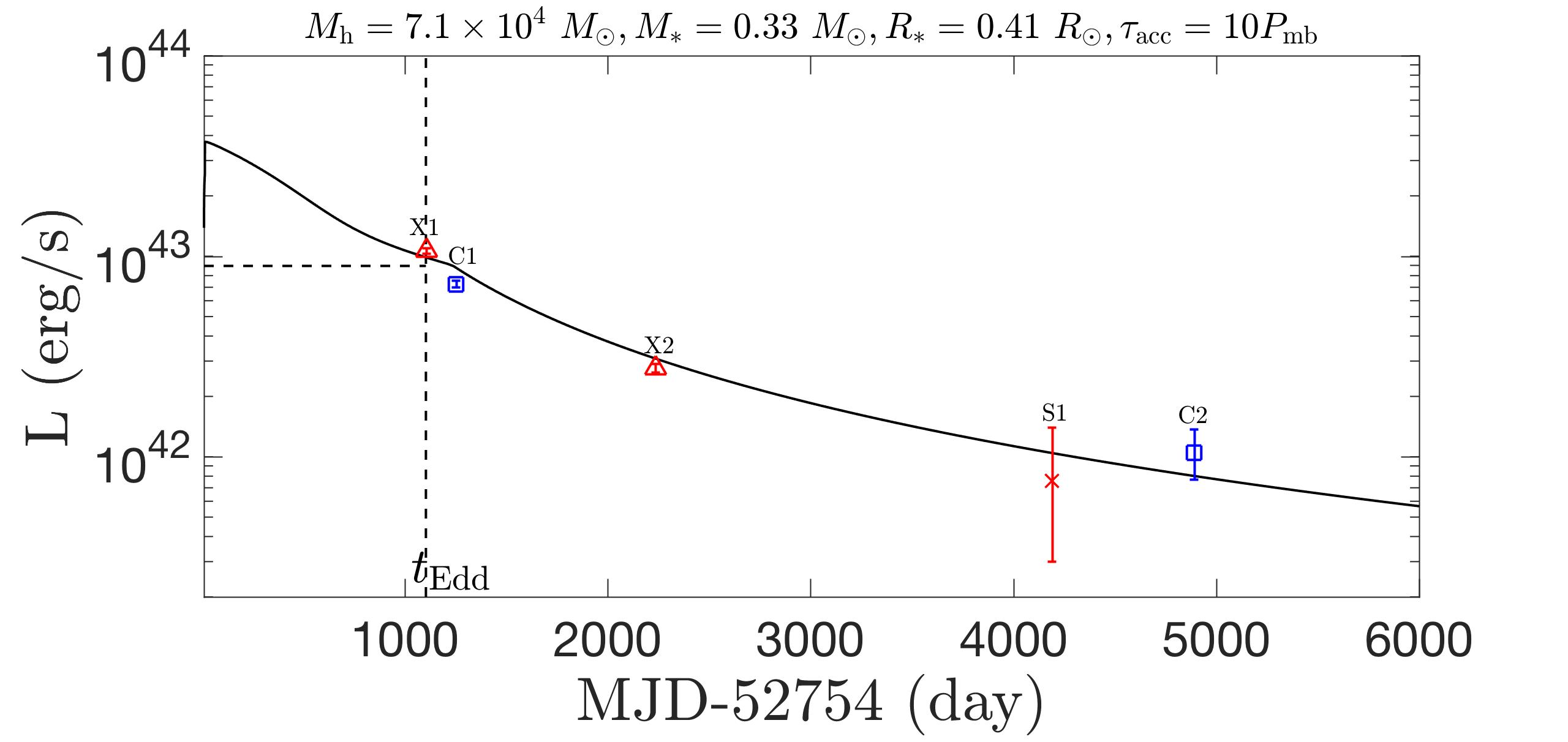}
\caption{Modeling of the bolometric luminosity evolution of J2150-0551 (solid line). The \textit{Chandra}, \textit{XMM-Newton}, and \textit{Swift} data points are from \cite{Lin_A_2018} and are shown as blue squares, red triangles, and a green cross, respectively, with 90\% error bars.}
\label{candidateL}
\end{figure}

After this paper was submitted, we noticed that \cite{Lin_A_2018} reported an IMBH-TDE candidate X-ray source, 3XMM J215022.4-055108 (hereafter J2150-0551), in a globular cluster in the galaxy 6dFGS gJ215022.2-055059. We model this source by assuming the X-ray luminosity data points C1, X2, S1, C2 from \cite{Lin_A_2018} to be in the sub-Eddington phase, which follow the $t^{-5/3}$ power-law decay. We choose an accretion timescale $\tau_{\rm acc} = 10P_{\rm mb}$ to fit the super-Eddington data point X1 and the subsequent luminosity decay. Figure \ref{candidateL} shows the fit. Thus, we obtain the approximate date of the disruption to be MJD 52754 and $t_{\rm Edd} \simeq 1104\ {\rm day}$. We obtain the Eddington luminosity $ \sim 0.9 \times 10^{43}\ {\rm erg\ s^{-1}}$, which corresponds to a BH mass $\sim 7.1 \times 10^4\ M_{\odot}$. We can infer the object to be a main-sequence star of $M_* \simeq 0.33\ M_{\odot}$ and $R_* \simeq 0.41\ R_{\odot}$, using the relation $R_* \propto M_*^{0.8}$.

We also model the evolution of the spectral temperature $T \sim T_{\rm eff} = [L/4\pi R_{\rm ph}^2 \sigma]^{1/4}$. Let $t_{\rm c}$ be the time when the wind photosphere recedes to the circularization radius $R_{\rm c}$ (also the outer radius of the disk). For early times, $t < t_{\rm c}$, $R_{\rm ph}$ is given by Eq. (\ref{Rph}), and so is $t_{\rm c}$. For $t \gtrsim t_{\rm c}$, the wind launching does not stop abruptly since $\dot{M}_{\rm acc}$ is still larger than, although close to, $\dot{M}_{\rm Edd}$. It is unclear how exactly $\dot{M}_{\rm w}(t)$ would diminish with time, but one can expect that the wind launching region on the disk would shrink inward. Thus, we assume $R_{\rm ph} \simeq R_{\rm c} (t/t_{\rm c})^{-x}$, where $x$ is a constant index whose value is to be determined. This is equivalent to saying that the inner boundary of the \textit{visible} outer disk region is advancing inward. This phase lasts until $R_{\rm ph}$ recedes all the way to the inner radius of the disk, i.e., the innermost stable circular orbit radius $R_{\rm isco}$. Thereafter, the emission would be entirely from a bare disk.

The spectral (color) temperature $T$ may be slightly higher than $T_{\rm eff}$ due to some spectral hardening effects, the most important of which is the Compton scattering on hot electrons in the disk atmosphere. This is usually accounted for by a temperature hardening factor in $T= f_{\rm c} T_{\rm eff}$ \citep{Shimura_On_1995,Merloni_On_2000,Sadowski_Slim_2011}. Here we take $f_{\rm c}= 1.3$. Our model prediction of $T(t)$ is shown in Figure 8, along with the spectral fit result for J2150-0551 from \cite{Lin_A_2018}. Here we find $t_{\rm c} \approx 505$ days, $f_{\rm w} = 2$, and $x \approx 5.4$, and we assume the BH has the maximum spin, so $R_{\rm isco} \simeq 0.5~R_{\rm S}$.    

\begin{figure}
\centering
 \includegraphics[scale=0.1]{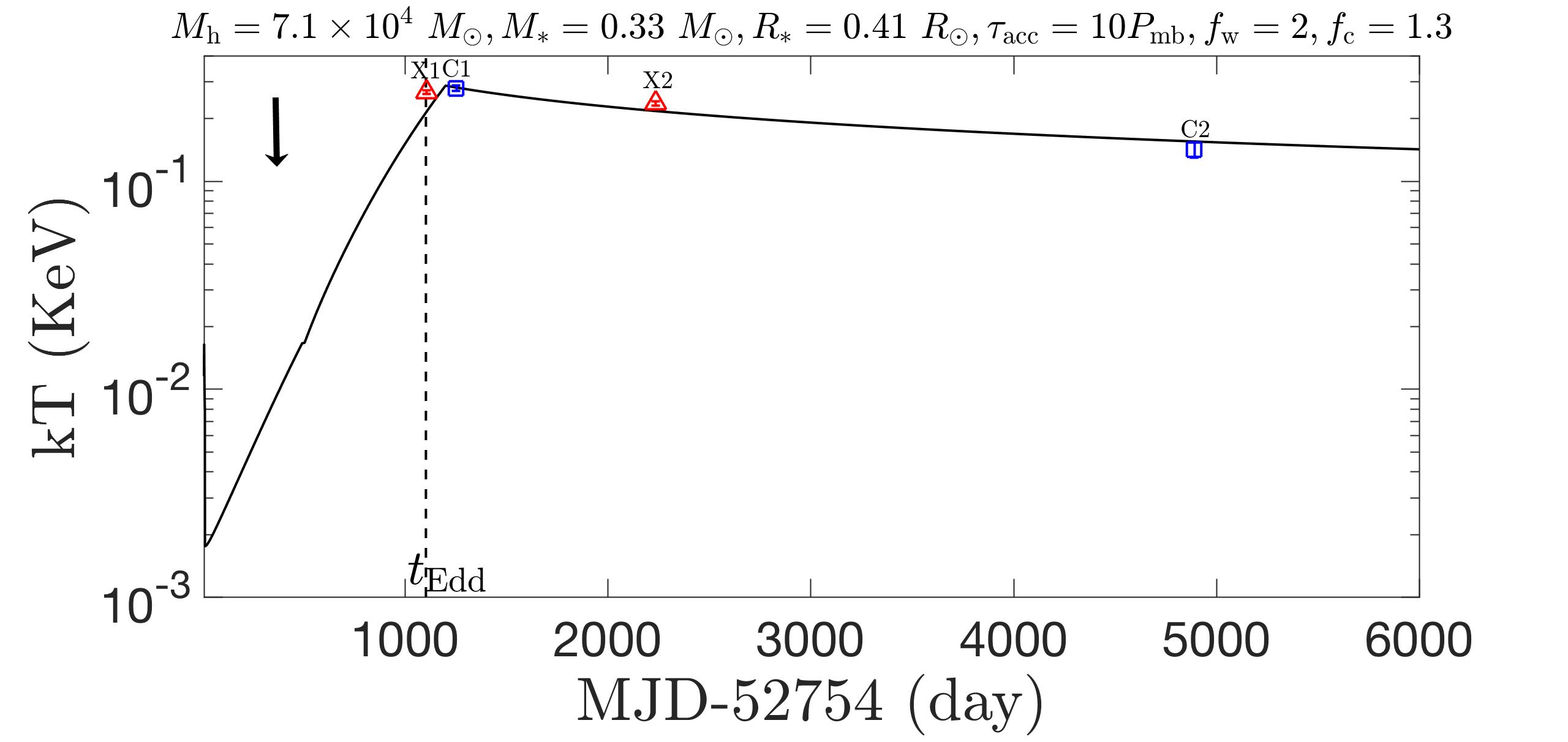}
\caption{Modeling of the temperature evolution of J2150-0551. Data points with 90\% error bars are from \cite{Lin_A_2018} and the arrow shows the time of the non-detection of the source in the \textit{XMM-Newton} slew observation on 2004 May 14.}
\label{candidateTem}
\end{figure}

\begin{figure}
\centering
 \includegraphics[scale=0.1]{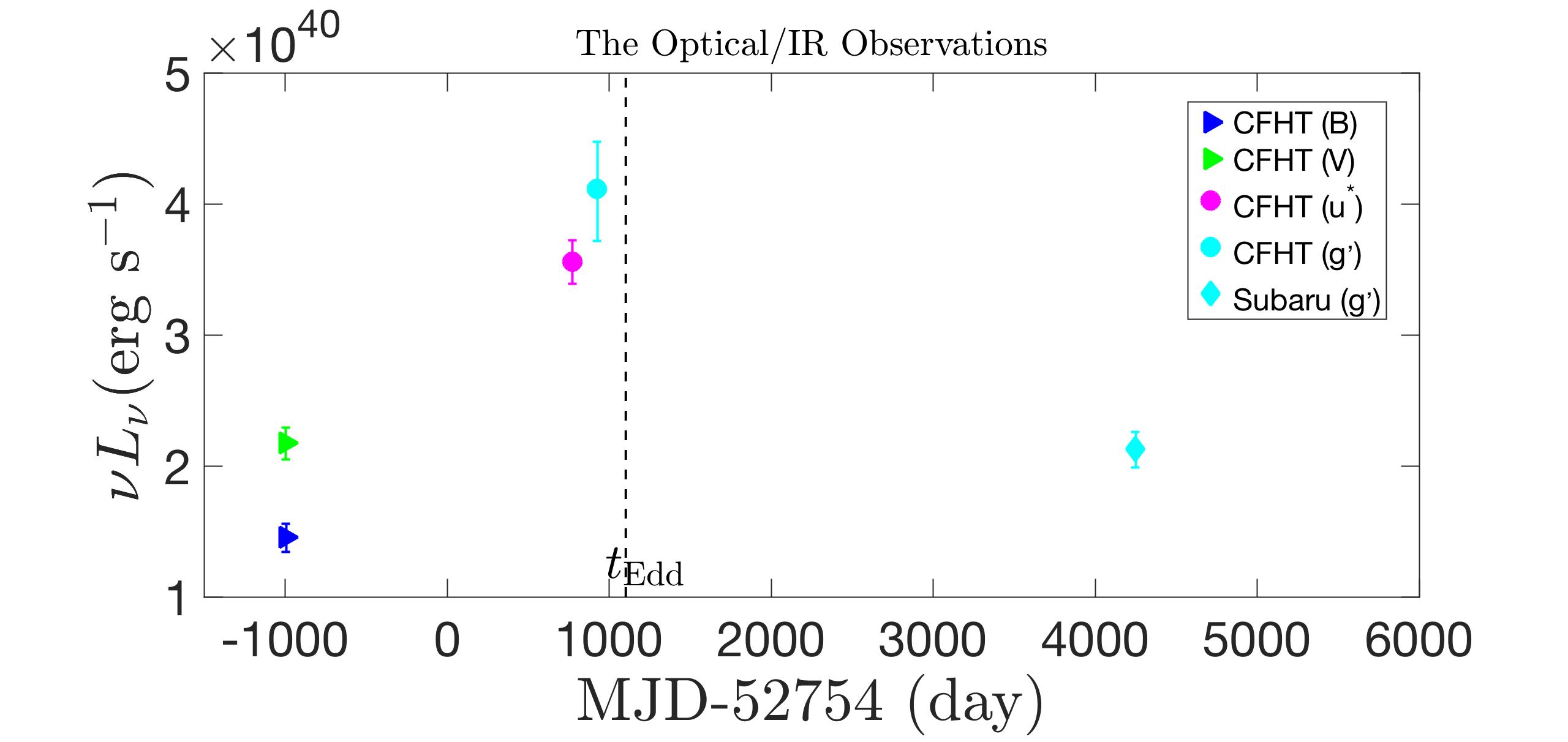}
\caption{Optical flare of J2150-0551. Data are taken from \cite{Lin_A_2018}, and we choose only those within the narrow wavelength range between 350 nm and 550 nm. }
\label{candidateO}
\end{figure}

In general, our model predictions are consistent with the data. We can conclude in Figure \ref{candidateL}-\ref{candidateTem} that most of the radiation is emitted in the UV and optical bands in the super-Eddington phase, and the X-ray emission appears later. This is roughly consistent with the detected optical flare (shown in Figure \ref{candidateO}) between 2005 May and November before the X-ray detections, and is also consistent with the non-detection of X-rays on 2004 May 14 (as the arrow indicates in Figure \ref{candidateTem}). 

As was pointed out in \cite{Lin_A_2018}, the temperature has a marginal rise from X1 to C1, even though C1 has already dimmed with respect to X1, which is unexpected for a bare, thermal disk. This behavior is qualitatively consistent with our model (see Figure \ref{candidateTem}), in which the wind launching after X1 is diminishing but has not yet died off.

\section{Summary and discussion}
\label{discussion}

The existence of the IMBH population is still a mystery. Nevertheless, the search for them and their subsequent identification has a great impact on the understanding of the seeds and growth history of supermassive BHs (SMBHs). We considered the disruption of a main-sequence star by an IMBH, focusing on the observational features in the aftermath. Due to a slow process of debris circularization characteristic of this type of disruption, we suggest that the mass accretion rate follows Eq. (\ref{accretion_rate_eq}). At the early times, the accretion rate is super-Eddington and produces a disk wind to regulate the luminosity. After the accretion rate dropps blow the Eddington limit, the wind ceases. As demonstrated in Figure~\ref{luminosity}, after a very rapid rise, the super-long-time radiative luminosity is continuously high at around the Eddington luminosity and lasts for more than 10 yr.

We also estimated the radius of the photosphere and the temperature, both of which are affected by the assumed accretion timescale. Low accretion with a larger accretion timescale will produce a smaller and hotter photosphere for much longer.  The spectrum peaks in the UV band until the photosphere recedes and then it moves to the far UV and soft X-ray.

Our estimate of the spectral property is based on the assumption of a quasi-spherical photosphere and blackbody spectrum. In reality the spectrum might be very complicated, and depend on the circumstances near the source and the line of sight \citep{Dai_A_2018}. Also the outflow could cool down and recombine so as to absorb and reprocess a fraction of the X-rays into the optical continuum and line emission \citep{Metzger_A_2016,Roth_THE_2016}. This process could change the shape of the spectrum.

Recently, \cite{Tremou_The_2018} searched for radio emission from a sample of GCs, assuming it to be a sign of accreting IMBHs. Further assuming that the empirical `fundamental plane' derived from those bright, well fed, stellar-mass or supermassive BH candidates also applies to starving IMBHs in GCs, the radio non-detections in GCs led those authors to conclude that IMBHs with masses $ \gtrsim 10^3 M_{\odot}$ are rare or absent in GCs. However, we found that a IMBH TDE will emit mainly in the UV/optical, and the kinetic luminosity of the wind (which is supposed to produce radio emission) is $\ll L_{\rm Edd}$. The gas near an IMBH in the GC might be very rare, so the radio emission due to shock interaction may be weak. Therefore, we propose that searching for TDEs in UV/optical bands will be a better method of finding IMBHs in GCs.

Recent work by \cite{Chilingarian_A_2018} estimated the masses of IMBHs in a sample of low-luminosity active galactic nuclei (AGNs) by analyzing the width and the flux of broad H$_{\alpha}$ emission lines. Whether some of these low-luminosity AGN-like events are actually IMBH TDEs is an interesting question. Answering it demands a long-term ($\gtrsim 10\ {\rm yr}$) but coarse optical monitoring to establish the luminosity and spectrum evolution trends reminiscent of what we predicted here.

Finally, we modeled the recently reported IMBH-TDE candidate J2150-0551, and found that the data are consistent with our predictions, including the optical flare in the super-Eddington phase and the X-ray emission in the sub-Eddington phase.

\section{acknowledgments}
We thank Tsvi Piran, Enrico Ramirez-Ruiz, and Nicholas Stone for helpful discussions, and thank the referee for helpful comments and suggestions. This work is supported by NSFC grant No. 11673078.

\begin{appendix}
\label{Appendix}
\section{Main effects on the efficiency of circularization}
Here we consider four main effects that dissipate the energy of the debris stream and circularize its orbit.
\subsection{Periastron nozzle shock}
As the stream approaches the pericenter, it is compressed like an effective nozzle. This effect will be accompanied by energy dissipation and redistribution of the angular momentum. In \cite{Guillochon_PS1_2014}, their hydrodynamical simulation of the debris stream evolution after the disruption of a main-sequence star by a $10^3M_{\odot}$ BH with a penetration parameter $\beta = 2$ shows that about 10\% of the kinetic energy of the stream is dissipated by the nozzle shock upon the pericenter passage. Although it is larger than the theoretical value of the fraction of orbital energy dissipated in the nozzle shock, which is $\sim \beta(M_{\rm h}/M_*)^{-2/3} = 2\%$ \citep{Kochanek_The_1994}, it is still smaller than the specific energy required to dissipate in order to circularize (Eq. \ref{circularization_energy}).

\subsection{Apsidal intersection}

\begin{figure}
\centering
 \includegraphics[scale=0.1]{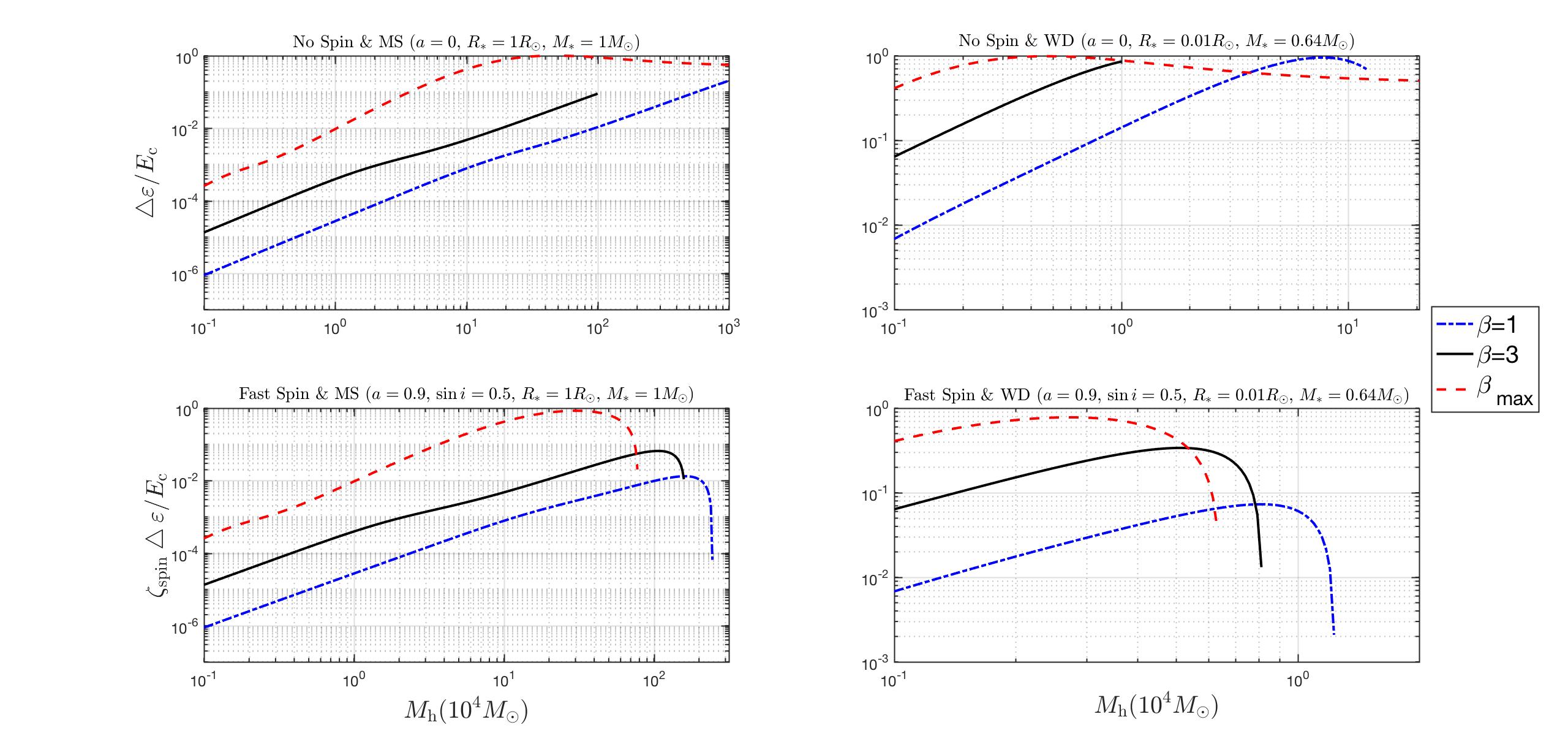}
\caption{Heating efficiency of GR apsidal intersection as a function of BH mass. The four figures represent different TDE situations, i.e. no or fast spin of a BH disrupting a main-sequence star or a white dwarf. The blue dot-dashed lines, black solid lines, and red dashed lines represent the penetration factor $\beta = 1$, $\beta = 3$ and upper limit, respectively. We calculate the efficiency if the disrupted star is a Sun-like star ($R_* = R_{\odot}$, $M_* = M_{\odot}$) or a typical white dwarf ($R_* = 0.01R_{\odot}$, $M_* =  0.64M_{\odot}$). If the spin of the BH is very fast (assuming the dimensionless spin of the BH, $a = 0.9$, and a moderate inclination of the star's orbit to the BH's spin plane, $\sin{i} = 0.5$, the Lense--Thirring effect would cause a collision offset due to the out-of-plane precession which would largely reduce the heating efficiency of the GR apsidal intersection. We consider this effect to be efficient if the efficiency is higher than $ \sim 0.1$, and is inefficient otherwise. We conclude that no matter whether the BH has spin or not, or how close the pericenter is  to it, tidal disruption of a main-sequence star by an IMBH should always be a slow accretion process, which is different from the TDEs for SMBHs.}
\label{GRdissipation}
\end{figure}

As the stellar streams begin to return to pericenter, the earlier returning streams will have larger apsidal angle of precession than the later ones, due to general relativistic (GR) apsidal precession \citep{Rees_Tidal_1988,Shiokawa_GENERAL_2015}. Then the outgoing stream will collide with the infalling stream near the apocenter \citep{Dai_Soft_2015, Jiang_PROMPT_2016}. We can estimate this effect quantitatively. Assuming a Schwarzschild IMBH, the precession angle after a single orbit is
\begin{eqnarray}
\phi &=& \frac{3 \pi}{2} \frac{R_{\rm S}}{R_{\rm p}} \nonumber
\\
&\simeq& 0.54\degree~\beta M_4^{2/3} r_*^{-1} m_*^{1/3}.
\end{eqnarray}
The radius of the intersection between the original ellipse and the shifted ellipse is
\begin{equation}
R_{\rm I} \simeq \frac{(1+e_{\rm mb})R_{\rm T}}{\beta(1-e_{\rm mb} \cos{(\phi /2)})}.
\end{equation}   
This is about $3.2 \times 10^{13}\ {\rm cm}$ for $10^4M_{\odot}$ IMBH with $\beta=1$. The intersection angle of the outgoing most bound orbit with the incoming stream is
\begin{equation}
\cos{\Theta} = \frac{1-2\cos{(\phi/2)}e_{\rm mb}+\cos{\phi}~e_{\rm mb}^2}{1-2\cos{(\phi/2)}e_{\rm mb}+e_{\rm mb}^2}.
\end{equation}
This is about $5\degree.4$ for a $10^4M_{\odot}$ IMBH with $\beta=1$. Here, we adopt an inelastic collision model to calculate the energy loss. Assuming the mass and velocity of two streams are similar, so the post-collision velocity of the stream at the intersection point is $v_{\rm f} = v_{\rm i} \cos{(\Theta/2)}$, where $v_{\rm i}$ is the velocity of the streams just before collision, given by $E_{\rm mb}= -GM_{\rm h}/R_I + v_{\rm i}^2/2$. Then the specific energy dissipation is
\begin{equation}
\bigtriangleup \varepsilon = GM_{\rm h} \left(\frac{1}{R_{\rm I}} - \frac{1}{2a_{\rm mb}}\right) \sin{(\Theta/2)}^2.
\end{equation}
For a $10^4M_{\odot}$ IMBH with $\beta = 1$, the energy dissipation upon apsidal intersection is about $10^{12}\ {\rm erg/g}$. This is negligible compared with $E_c$ (Eq. \ref{circularization_energy}) about IMBH. But it becomes important as BH mass increases, as is shown in Figure \ref{GRdissipation}, which is roughly consistent with the simulation results of \cite{Hayasaki_Finite_2013}. Furthermore, the efficiency of circularization is sensitive to BH spin. The Lense--Thirring effect would induce deflections out of the original orbital plane that cause the collision offset $\zeta_{\rm spin}$ from the apsidal intersection, and reducing the heating efficiency of the GR apsidal intersection \citep{Cannizzo_The_1990,Guillochon_A_2015,Hayasaki_Circularization_2016,Jiang_PROMPT_2016}.

\subsection{Thermal viscous shear}

\begin{figure}
\centering
 \includegraphics[scale=0.1]{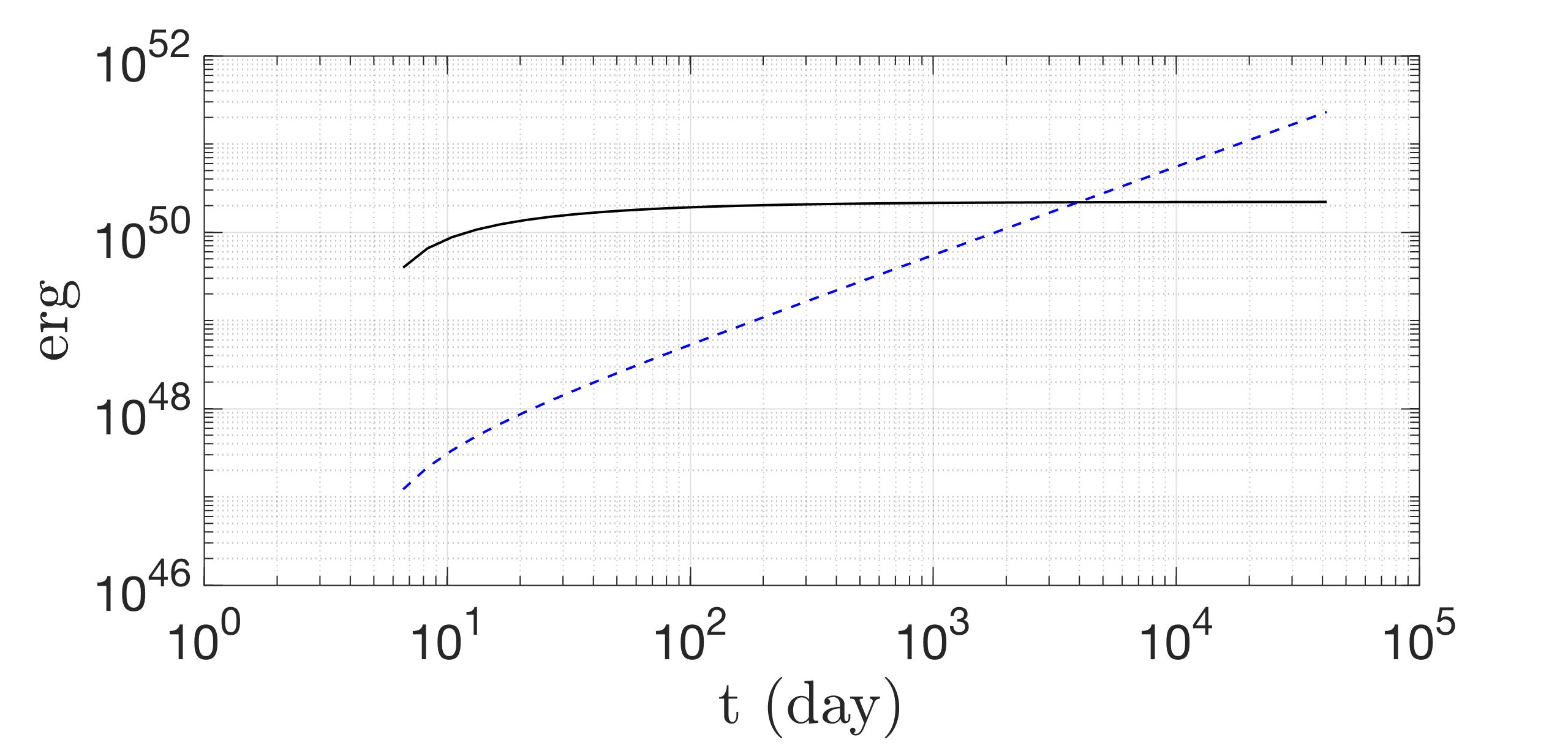}
\caption{Circularization history by thermal viscous shear. After an IMBH ($M_{\rm h} = 10^4 M_{\odot}$) tidally disrupts a Sun-like star, the fresh stellar material will continuously be supplied to the eccentric disk, producing a large amount of the required dissipated energy for circularization, as the black solid line shows. The blue dashed line represents the thermal viscous heating history. Initially the thermal viscous heating energy is far below the required dissipated energy for circularization until the process of circularization is complete at $\sim 4000$ days. So this effect is very inefficient at circularizing the stellar debris.}
\label{thermal_viscous}
\end{figure}

As the star is disrupted, its debris has similar specific angular momentum $j_*  \simeq (2G M_{\rm h} R_{\rm p})^{1/2}$, and this would remain for a long time until the shear forces redistributed its angular momentum. After pericenter passage, the star would expand adiabatically, the stellar stream at neighboring radii would move with different angular velocity, and would suffer from a viscous torques. Despite the lack of detail of the streams' dynamics, we can estimate the upper limit of the dissipated energy converted from the orbital kinetic energy. The heating rate per unit area is
\begin{equation}
\label{heating_rate}
Q_{\rm vis}^+ = \int_{-\infty}^{+\infty} \nu_{\rm vis} \rho \left(R\frac{d\Omega}{dR}\right)^2\ dz,
\end{equation}
where $\Omega = j_*/R^2$ is the angular velocity of the stream. The kinematic viscous shear is
\begin{eqnarray}
\label{kinematic_viscous}
\nu_{\rm vis} &=& -\alpha c_s^2 \left(R\frac{d\Omega}{dR}\right)^{-1} \nonumber
\\
&\simeq& -\alpha \frac{P}{\rho} \left(R\frac{d\Omega}{dR}\right)^{-1},
\end{eqnarray}
Here $c_s$ is the sound speed and $\alpha \sim 0.1$ is the viscous parameter. If we assume the height-to-radius ratio $H/R \sim 1$ and substitute Eq. (\ref{kinematic_viscous}) into Eq. (\ref{heating_rate}), the integration becomes
\begin{equation}
Q_{\rm vis}^+ = \int_{0}^{R} 2\alpha P \frac{j_*}{r^2}~dz.
\end{equation}
We assume the radiation pressure dominates the total pressure and let
\begin{equation}
P \sim \frac{G M_{\rm h}}{\kappa_{\rm es} R^2}
\end{equation}
be an upper limit, where $\kappa_{\rm es} \simeq 0.34~{\rm cm^2\ g^{-1}}$ is the electron scattering opacity. We then have
\begin{equation}
Q_{\rm vis}^+ = 4 \alpha \frac{GM_{\rm h} j_*}{\kappa_{\rm es} R^3}.
\end{equation}
So the heating rate for thermal viscous shear is
\begin{eqnarray}
L_{\rm vis} &\simeq& \int_{R_{\rm in}}^{R_{\rm c}} Q_{\rm vis}^+ 2 \pi R~dR \nonumber
\\
&\simeq& 8 \pi \alpha \frac{G M_{\rm h} j_*}{\kappa_{\rm es} R_{\rm in}} \nonumber
\\
&\simeq& 3.2 \times 10^{42}\ \alpha_{-1} \beta^{-1/2} M_4^{2/3} r_*^{1/2} m_*^{-1/6}\ {\rm erg/s},
\end{eqnarray}
where $R_{\rm in} = 1.5R_{\rm S}$ is the last stable orbit for a Schwarzschild BH.
 
As the fresh debris returns to the pericenter, we suggest that the thermal viscous heating has begun and most of the heat comes from the inner region near $R_{\rm in}$. As illustrated in Figure \ref{thermal_viscous}, when the total thermal viscous heating energy is greater than the required dissipated energy for circularization
\begin{equation}
L_{\rm vis} (t-P_{\rm mb}) \gtrsim \int_{P_{\rm mb}}^{t} E_c \dot M_{\rm fb}~dt',
\end{equation}
we can say that the circularization is complete. Although $L_{\rm vis}$ is quite large ($\sim 10L_{\rm Edd}$), the circularization is still very inefficient due to the large amount of material supplement at the beginning of the TDE. For this case (a Sun-like star disrupted by a $10^4 M_{\odot}$ IMBH), the end time of circularization is about 4000 days, as shown in Figure \ref{thermal_viscous}.

\subsection{Magnetic shear}
\label{magnetic}
The magnetic field in the stellar debris might be amplified by MRI as the streams travel along its elliptical orbits. \cite{Bonnerot_Long_2017} suggest magnetic stress could hasten circularization. We estimate the ratio of Alfv$\acute{e}$n velocity and circular velocity at the apocenter $v_{\rm A}/ v_{\rm c} \sim 10^{-6}$ with a seed field strength $\sim 1$ G for Sun-like star, where $v_{\rm A}^2 \simeq B^2/(4 \pi \rho)$ and the density of the stream $\rho \sim \dot M_{\rm peak}/(\pi R_{\odot}^{2} v_{\rm c}) \sim 10^{-4}\ \rm {g/cm^3}$. So the magnetic stress can be neglected for the full disruption of a Sun-like star unless the MRI could amplify the magnetic energy at least by four orders of magnitude in the early stream evolution, although this is unlikely because the MRI does not have time to reach saturation \citep{Stone_Three_1996}. \cite{Chan_Magnetorotational_2018} studied how linear evolution of MRI in an eccentric disk amplifies magnetohydrodynamic perturbations, but how the magnetic stress affects the debris is unclear. 

In summary, these effects, including periastron nozzle shock, apsidal intersection, thermal viscous shear, and magnetic shear, are all very inefficient for circularization. GR apsidal intersection has long been thought to be a main factor for dissipation in SMBH TDEs. Thus the accretion process would be very rapid for a TDE with an SMBH of low spin \citep{Hayasaki_Finite_2013}. However, for the disruption of a main-sequence star by an IMBH, this effect is very inefficient. So the debris will remain around the IMBH for a long time in this case. In Figure \ref{GRdissipation}, we considered four parameters: the BH spin, the BH mass, the penetration factor, and the type of the disrupted star. The heating efficiency of GR apsidal intersection would increase with the mass of the BH and the penetration factor. However, the Lense--Thirring effect would cause the collision offset due to the out-of-plane precession, which would largely reduce its heating efficiency. We consider this effect to be efficient if the efficiency is higher than $\sim 0.1$ and inefficient otherwise. We can see in Figure \ref{GRdissipation} that no matter whether the BH has spin or not, or how close the pericenter is to it, tidal disruption of a main-sequence star by an IMBH is always a long-term circularization process.

\end{appendix}

\bibliographystyle{yahapj}
\bibliography{cited}

\end{document}